\documentclass[twocolumn]{aastex631}
\usepackage{graphicx} 
\usepackage[utf8]{inputenc}
\usepackage{indentfirst}
\usepackage{color,soul}

\usepackage[hyphens]{url}
\usepackage{amsmath}
\usepackage{amssymb}
\usepackage{bm}
\usepackage{csquotes}
\MakeOuterQuote{"}
\usepackage{hyperref}
\hypersetup{breaklinks=true}
\usepackage{makecell}
\usepackage{lipsum,booktabs}

\newcommand{\Msun}{\,\mathrm{M}_\odot}
\newcommand{\Zsun}{\,\mathrm{Z}_\odot}

\begin{document}

\title{Beyond the Goldilocks Zone: Identifying Critical Features in Massive Black Hole Formation}

\author[0000-0002-4100-4521]{Elizabeth Mone}
\affiliation{Center for Relativistic Astrophysics, School of Physics, Georgia Institute of Technology, \\
837 State Street, Atlanta, GA 30332, USA}

\author[0000-0003-4811-9863]{Brandon Pries}
\affiliation{Center for Relativistic Astrophysics, School of Physics, Georgia Institute of Technology, \\
837 State Street, Atlanta, GA 30332, USA}

\author[0000-0003-1173-8847]{John H.\ Wise}
\affiliation{Center for Relativistic Astrophysics, School of Physics, Georgia Institute of Technology, \\
837 State Street, Atlanta, GA 30332, USA}

\author{Sandrine Ferrans}
\affiliation{Center for Relativistic Astrophysics, School of Physics, Georgia Institute of Technology, \\
837 State Street, Atlanta, GA 30332, USA}

\correspondingauthor{emone3@gatech.edu}


\begin{abstract}
Most galaxies, including the Milky Way, host a supermassive black hole (SMBH) at the center. These SMBHs can be observed out to high redshifts ($z\geq$ 6) if the accretion rate is sufficiently large. However, we do not fully understand the mechanism through which these black holes form at early times. The heavy (or direct collapse) seeding mechanism has emerged as a probable contender in which the core of an atomic cooling halo directly collapses into a dense stellar cluster that could host supermassive stars that proceed to form a BH seed of mass $\sim\!10^{5} \Msun$. We use the \textit{Renaissance} simulations to investigate the properties of 35 DCBH candidate host halos at $z = 15-24$ and compare them to non-candidate halos. We aim to understand what features differentiate halos capable of hosting a DCBH from the general halo population with the use of statistical analysis and machine learning methods. We examine 18 halo, central, and environmental properties. We find that DCBH candidacy is more dependent on a halo's core internal properties than on exterior factors such as Lyman-Werner flux and distance to closest galaxy; our analysis selects density and radial mass influx as the most important features (outside candidacy establishing features). Our results concur with the recent suggestion that DCBH host halos neither need to lie within a "Goldilocks zone" nor have a significant amount of Lyman-Werner flux to suppress cooling. This paper presents insight to the dynamics possibly occurring in potential DCBH host halos and seeks to provide guidance to DCBH subgrid formation models.
\end{abstract}


\section{Introduction} \label{sec:intro}
Supermassive black holes (SMBHs) in the early universe remain a topic of intense study in the astronomical community. SMBHs have been observed in distant galaxies out to redshifts approaching $11$ \citep[e.g.,][]{Cameron2023,Goulding2023,Abuter2024,Bogdan2024,Bosman2024,Kovacs2024,Maiolino2024a,Maiolino2024b,Marshall2024,Natarajan2024,Scholtz2024}. However, their formation mechanism is still unclear.


There are three primary seeding mechanisms leading to the formation of black holes \citep{Volonteri2010a,Inayoshi2020,Sassano2021,Ellis2024,Regan2024}. First is a light seed ($M_{\bullet} \lesssim 10^{2} \Msun$) that forms after a massive star goes supernova. The most probable source are metal-free stars (Population III; Pop III), whose initial mass function is thought be top-heavy \citep{Klessen2023}. Most of these seeds do not exist at the center of massive halos and have a limited gas supply because of intrinsic and nearby stellar feedback \citep{Smith2015}. This seeding mechanism is thus unlikely to produce a SMBH in the early stages of the universe \citep{Regan2020a}; however, it may be possible for a small fraction of light seeds \citep{Lupi2024a,Lupi2024b,Mehta2024}. Second are intermediate mass seeds ($M_{\bullet} \sim\! 10^{3} \Msun$) which form from stellar collisions that precede black holes \citep[e.g.,][]{Sakurai2017,Reinoso2018,Sakurai2019,Regan2020c,GonzalezPrieto2024,Vergara2024,Woods2024}. Lastly is the heavy seed or direct collapse mechanism ($M_{\bullet} \gtrsim 10^{4} \Msun$) \citep{Latif2013,Agarwal2016a,Suazo2019,Chon2020}, which is the mechanism we will explore within this paper. Direct collapse black holes (DCBHs) can form in pre-galactic atomic cooling halos with low metallicities and low initial concentrations of molecular hydrogen $\left(\rm{H}_{2}\right)$, the dominant coolants in typical star-forming clouds \citep{Bromm2003,Begelman2006,Begelman2008,Dijkstra2008,Volonteri2010a,Volonteri2010b}. In the absence of efficient coolants, the collapsing halo fails to rigorously fragment and form stars leading to a buildup of gas in a single Jeans-unstable cloud of $M \sim\! 10^{4}-10^{5} \Msun$ forming supermassive stars above $1000 \Msun$ and perhaps up to $10^{5} \Msun$ \citep{Regan2020c,Volonteri2021,Patrick2023,Prole2024}. At these masses, the core becomes unstable under general relativity and collapses directly into a black hole of $\sim\! 10-20 \Msun$, which then accretes the surrounding envelope and grows to $\sim\! 10^{3}-10^{4} \Msun$ \citep{Begelman2006,Volonteri2012,Zwick2023,Mayer2024}. This heavy seed formation mechanism forgoes the issue of super-/hyper-Eddington accretion that would be required in light seeds for them to reach supermassive scales after only a few hundred million years. DCBH candidates have been identified observationally \citep{Agarwal2016b,Pacucci2016,Larson2023,Nabizadeh2024,Natarajan2024}, though further observations are necessary to attempt to determine their seeding mechanism.

Early efforts to estimate DCBH detectability focused on the estimated luminosity in one or more of the planned JWST bands \citep[e.g.,][]{Volonteri2010b}. Initial candidacy detections employed different methods: \citet{Pacucci2016} used multiwavelength photometry selections, targeting candidates bright in both the IR and X-ray regimes; \citet{Agarwal2016a} used spectroscopic observations combined with source morphology to argue that their source of interest was a candidate due to LW radiation from neighboring stellar clusters. Most modern observations rely on either multiwavelength and/or spectroscopic observations; IR observations are typically required to confirm the high redshift nature of these objects, and most candidacy selections invoke an X-ray detection to distinguish it from stellar systems \citep[e.g.,][]{Kovacs2024,Natarajan2024}. Some also compare observed spectra to theoretical predictions \citep{Nabizadeh2024,Natarajan2024}.

Atomic cooling halos are characterized by virial temperatures of $T_{\rm{vir}} \gtrsim 10^{4} \, \rm{K}$, corresponding to $\sim\! 3 \times 10^{7} \Msun$ at $z = 15$. In this regime, halos can efficiently cool via atomic hydrogen to $8000 \, \rm{K}$ and collapse, as opposed to efficient $\rm{H}_{2}$ cooling down to $300 \, \rm{K}$ leading to greater fragmentation \citep{Bromm2003,Yoshida2003}. Molecular hydrogen cooling can be suppressed by Lyman-Werner (LW) radiation, in a band of $11.2-13.6 \, \rm{eV}$, that dissociates $\rm{H}_{2}$. The universe is optically thin in the LW band \citep{Haiman1997a,Haiman1997b,Holzbauer2012}, so the high-redshift galaxy population builds up the LW background that can effectively suppress primordial star formation. Starlight from galaxies within a few proper kiloparsecs can dominate over the background that motivates the "close pair" scenario of DCBH formation \citep{Dijkstra2008}. However during the collapse, the $\rm{H}_{2}$ in the core becomes self-shielded from external LW radiation from the increasing $\rm{H}_{2}$ column density \citep{Wolcott-Green2011,Patrick2023}. The core cools through this reaction to $\sim 500 \, \rm{K}$, accelerating the collapse and enhancing fragmentation but not to the point of present-day star formation \citep{Wise2019}. $\rm{H}_{2}$ cooling can also be suppressed through alternative mechanisms such as rapid halo growth and strong shocks from cold accretion flows \citep{Fernandez2014,Wise2019,Kiyuna2024}.

The main objective of this work is to assess the importance of halo, central, and environmental features in potential DCBH host halos. To accomplish this we make use of zoom-in cosmological simulations of the first stars and galaxies to evaluate statistical differences between halos selected as candidates of the direct collapse seeding mechanism and all other atomic cooling halos. Additionally, we provide a statistical and supervised machine learning based ranking of features important in selecting candidate DCBH halos. The purpose of this study is to understand the necessary conditions for a DCBH to form.

The structure of the paper is outlined as follows. In Section~\ref{sec:methods}, we discuss the data used in this analysis, as well as the statistical analysis and feature selection methods used to analyze our data. In Section~\ref{sec:res}, we present our results. In Section~\ref{sec:dis}, we discuss the applications and implications of our results. In Section~\ref{sec:con}, we conclude and summarize our findings.

\section{Methods} \label{sec:methods}
\subsection{Renaissance Simulations} \label{sec:renaissance}
The \textit{Renaissance} Simulations were run on the NCSA Blue Waters machine at NCSA with \texttt{\textsc{Enzo}}, an adaptive mesh refinement (AMR) code \citep{Bryan2014,Brummel-Smith2019}. These simulations model radiation transport of ionizing photons with ray tracing \citep{Wise2011} and $\rm{H}_{2}$ photodissociation from a background and from point sources, both of which are important in our analysis \citep{Regan2020a}. The LW background model is described in \citet{Xu2016}. The \textit{Renaissance} Simulations were first run with only dark matter (DM) to $z \sim\! 6$, when three regions of average, high, and low density were identified and named as the Normal, Rarepeak, and Void regions, respectively. These regions were resimulated with gas, star formation, and stellar feedback at higher mass and spatial resolutions and volumes ranging from $200-430 \, \rm{{Mpc}^{3}}$.

In this paper we analyze the Rarepeak region as more DCBH host candidates follows from the higher density \citep{Regan2020a}. The Rarepeak region encloses a volume of $(3.8 \times 5.4 \times 6.6) \, \rm{{Mpc}^{3}}$ with an effective resolution of $4096^{3}$ centering on two halos at $z = 6$ each with masses of $3 \times 10^{10} \Msun$. The resimulations of the Rarepeak region have a maximum spatial resolution of $19$ comoving $\rm{pc}$ with a dark matter particle mass resolution of $2.9 \times 10^{4} \Msun$. At this mass resolution, star formation is captured well below the atomic cooling limit while the parsec-scale resolution allows for gravitational collapse of the halo to be captured and analyzed for possible DCBH formation.

The \textit{Renaissance} Simulations do not have the ability to track if a gravitationally collapsing halo will form a DCBH, so instead we select halos that follow a specific set of criteria to be candidates for DCBH hosting. The criteria are as follows: starless halos at the atomic cooling limit ($T_{\rm{vir}} \sim\! 8000 \, \rm{K}$ and $M_{\rm{vir}} \sim\! 3 \times 10^{7} \Msun$ at $z = 15$) with low metallicity ($Z \lesssim 10^{-4} \Zsun$; see \citet{Chon2024}). We adopt the same relation as \citet{Fernandez2014} between virial mass and temperature of

\begin{equation}
    \label{eqn:acl}
    T_{\rm{vir}} = 0.75 \times 1800 \left(\frac{M_{\rm{vir}}}{10^{6} \Msun}\right)^{2/3} \, \left(\frac{1+z}{21}\right) \, \rm{K}.
\end{equation}

\noindent The Rarepeak simulation ends at $z = 15$; we analyze all atomic cooling halos from all 40 outputs that begin at $z = 24$ continuing down to the final redshift. Beyond our work, recent simulation suites, such as \texttt{BRAHMA} \citep{Bhowmick2024a} which applies a high resolution seeding model, are currently being used in a similar manner to constrain seeding mechanisms.

\subsection{Environment and Halo Variables} \label{subsec:var}
\citet{Regan2020a} identified a total of 76 DCBH candidates in the Rarepeak region whereas the Normal region contains three and the Void contains none. They studied the growth rates, LW fluxes, and distance to the nearest galaxy of these candidates. Their first conclusion was that while many candidates exhibited high growth rates this did not exclude non-candidates as many star-forming halos also exhibit fast growth. Additionally, they found the LW flux that most candidates experienced did not reach that necessary to suppress $\rm{H}_{2}$ formation and cooling, instead dynamical heating may play an important role in this suppression \citep[e.g.][]{Fernandez2014, Wise2019}. Lastly, while proximity to other galaxies provided increased LW flux, especially in the case of synchronized pair halos, it also increased the chances of a halo becoming enriched by metals; the conclusion was that a $10-100 \, \rm{kpc}$ range was the ideal distance for candidate halos \citep{Regan2020a}. In this paper, we extend upon the conclusions that \citet{Regan2020a} made by increasing the number of features considered as well as expanding upon comparisons made to non-DCBH candidate halos through the use of machine learning and statistical analysis. Other recent works such as \citet{Bhowmick2022,Bhowmick2024b} suggest a need for a comprehensive review of DCBH seeding requirements and mechanisms.

In the Rarepeak region, the candidates are chosen from atomic cooling halos to be starless and metal-free (average metallicity below critical metallicity of $10^{-4} \Zsun$). In order to prevent halos being double counted, those within the same formation lineage are removed. Non-candidate halos are selected to be within the same halo mass range ($M_{\rm{vir}} \sim\! 3.6 \times 10^{7} - 7.6 \times 10^{7} \Msun$) as the candidate halos. Additionally, a cut was made on non-candidate halos to remove those containing metal enriched stars younger than $20 \, \rm{Myr}$ to reduce peaks in LW flux from interior stars, as being starless is one of our requirements for direct collapse. This restricts the LW flux to external sources. With these criteria, our sample includes 35 DCBH candidate halos and $\sim\! 4000$ non-candidates.

The environment and halo properties that we consider can be separated into three categories and are as follows:
\begin{enumerate}
    \itemsep 0em
    \item \textbf{Central Halo Properties}: $\rm{H}_{2}$ fraction, Lyman-Werner Flux, Radial Mass Flux, Radial Velocity, Temperature, Density, Tangential Velocity, Turbulent Velocity
    \item \textbf{Halo Properties}: Tidal Field $t_{1}$ Eigenvalue, Gas Spin Parameter, Dark Matter Spin Parameter, Growth Rate, Mass, Metallicity, Stellar Mass
    \item \textbf{Environmental Properties}: Large-scale Overdensity, Distance to Closest Galaxy
\end{enumerate}

We include halo mass, stellar mass, and metallicity in our analysis to check our statistical analysis against the specified input parameters for DCBH candidates as outlined in Section~\ref{sec:renaissance}.

\subsubsection{Central Halo Properties} \label{subsub:chp}
In this paper many variables are calculated through the use of \texttt{yt} \citep{Turk2011}, the \texttt{\textsc{ROCKSTAR}} \citep{Behroozi2013} halo catalog, and the associated merger tree calculated with consistent-trees evaluated with \texttt{ytree} \citep{Smith2019}. To calculate the central properties of a halo we take a $50 \, \rm{pc}$ sphere centered on the halo's center of mass and calculate the mass-weighted average value.

The LW flux is calculated using the equation

\begin{equation}
    F_{\rm{LW}} = \frac{E_{\rm{LW}} k_{\rm{diss}}}{4\pi^{2} \sigma_{\rm{H2}} \nu_{\rm{H}}},
    \label{eq:LW_flux}
\end{equation}
where $E_{\rm{LW}}$ is the central LW radiation intensity ($12.4 \, \rm{eV}$), $k_{\rm{diss}}$ is the $\rm{H}_{2}$ dissociation rate taken from the radiation fields in the simulation outputs, $\sigma_{\rm{H2}} = 3.71 \times 10^{-18} \, \rm{{cm}^{2}}$ is the average effective $\rm{H}_{2}$ dissociation cross section in the LW band \citep[e.g.,][]{Wise2011}, and $\nu_{\rm{H}}$ is the Rydberg constant of hydrogen in Hertz.

The turbulent velocity is calculated by finding the standard deviation of the velocity corrected for the halo bulk velocity. The spherically-averaged radial gas mass flux is calculated with 

\begin{equation} \label{eq:rmf}
    \dot{m} = -4\pi r^{2} \rho v_{r},
\end{equation}
where $r$ is the radius, $\rho$ is the density, and $v_{r}$ is the radial velocity.

\subsubsection{Halo Properties} \label{subsub:hp}

Halo properties such as halo mass, stellar mass, and metallicity are calculated using a sphere out to the $R_{200}$ virial radius of the halo, containing an average overdensity of 200 times the critical density. The halo mass is the sum of the baryonic and dark matter mass, and stellar mass is the sum of all star particle masses within this sphere. We calculate metallicity as the mass-averaged value in the sphere. We considered both the baryonic and dark matter spin parameter, using the equation

\begin{equation}
    \lambda = \frac{J}{\sqrt{2} MVR},
    \label{eq:spin_param}
\end{equation}
where $J$ is total angular momentum, $M$ is the halo mass, $R$ is halo radius, and $V^{2} = GM/R$ \citep[e.g.][]{Bullock2001}.

The growth rate of the halo is calculated from the halo progenitor line dark matter mass both with respect to time and redshift. We apply \texttt{SciPy}'s \citep{Virtanen2020} Savitzky-Golay filter to calculate a smoothed derivative over a scale comparable to that of the average output time/redshift difference per halo. The parameters for the function are as follows: the \texttt{window\_length} (number of coefficients) is set to be equivalent to the length of the mass array; \texttt{polyorder} (polynomial order) has a maximum value of 3, being less when the number of mass points is small; \texttt{deriv} (order of the derivative) is set to first order; and \texttt{delta} (sample spacing) is proportional to the average timestamp differences for the halo, all other parameters are default. Due to the fact that the mass is taken from \texttt{ytree} which only follows dark matter it must be corrected for its gaseous component. We take this correction factor to be the cosmic fraction $\Omega_{m}/\left(\Omega_{m} - \Omega_{b}\right)$ where $\Omega_{m} = 0.266$ is the matter density parameter and $\Omega_{b} = 0.0449$ is the baryon density parameter that are used in the \textit{Renaissance} Simulations.

We calculate the large-scale tidal field of a halo using the process presented in \citet{Dalal2008} and \citet{DiMatteo2017}. The tidal force is best represented by the eigenvalues of the tidal tensor

\begin{equation}
    T_{ij} = S_{ij} - \frac{1}{3} \sum\limits_{i} {S_{ii}},
    \label{eq:tidal_tensor}
\end{equation}

\noindent where the strain tensor is given by $S_{ij} = \nabla_{i} \nabla_{j} \phi$ and $\phi$ is the potential. We calculate it in Fourier space by $\hat{S}_{ij} = k_{i} k_{j}/k^{2} \, \hat{\delta}$, where $\hat{\delta}$ is the density in Fourier space and $k$ is the wave number \citep{Dalal2008}. Once calculated in Fourier space, we apply an inverse transform to calculate the tidal tensor. From this we can find the eigenvalues $t_{1}$ the largest positive value and the direction in which the halo is stretched, $t_{2}$, and $t_{3}$ the largest negative value and direction in which the halo is compressed. By definition $t_{1} + t_{2} + t_{3} = 0$. The value of $t_{1}$ determines the halo's shape, where halos with high values of $t_{1}$ will form into disks due to higher angular momentum and low $t_{1}$ values lead to a spheroid shape. In \citet{DiMatteo2017} they determine a spheroid shape (low $t_{1}$) as ideal for growing high-redshift SMBHs.

\subsubsection{Halo Environment Properties}

To calculate the overdensity of the halo's environment we use a region of $15 \, \rm{kpc}$, which is approximately the turnaround radius of a halo of mass $3 \times 10^{10} \Msun$ at $z=8$, to assess whether DCBH host halos form in dense or sparse regions. The overdensity is calculated using $\delta(x) = \rho/\bar{\rho}$ where $\rho$ is the average density in the region and $\bar{\rho}$ is the critical density at the redshift of the halo.

To find the closest galaxy, \texttt{\textsc{ROCKSTAR}} \citep{Behroozi2013} was used to locate all the halos and associated galaxies at each redshift. We compare each of our halo positions to the positions of all the galaxies in the dataset, and identify the shortest distance with the constraints that the galaxy is contained within a well-resolved halo of a minimum mass of $1 \times 10^{7} \, \Msun$ and is not within the radius of the halo, i.e. is not a subhalo.

\subsection{Statistical Analysis} \label{subsec:stat}
With the data obtained through the calculations presented in Section~\ref{subsec:var} we perform statistical analysis to preliminarily form an understanding of which properties are most important in DCBH host candidacy. There are many statistical methods that can be used to accomplish this including basic \rm{Z}-score analysis, \rm{p}-value, and Mahalanobis distance. These statistical methods can be used as basic comparison between candidates (sample) and non-candidates (population), and they can also aid in performing feature selection.

\subsection{Feature Selection} \label{subsec:feature}
The results found by this paper aim to select features important in DCBH candidacy for future use by a support vector machine (SVM; see \citet{Grace2020} for an application to Pop III star formation). Feature selection is an essential process to reduce data dimensionality and confirm that results agree with physical interpretations \citep{Khalid2014}. Feature selection methods have been applied to other areas of astrophysics as well \citep[e.g.,][]{Montero2010,Donalek2013,Hoyle2015,Dhiman2018}. Along with statistical methods such as the Mahalanobis distance we can use supervised machine learning (ML) feature selection methods provided by \texttt{SciKit-Learn} \citep[\texttt{sklearn}]{Pedregosa2011}. These methods are however susceptible to correlation between variables, but a solution to this is provided in Section~\ref{subsubsec:perm}.

\subsubsection{Mahalanobis Distance} \label{subsub:mahalmethod}
The Mahalanobis distance is a multivariate expansion of the \rm{Z}-score which is defined as

\begin{equation}
    \rm{Z} = \frac{x-\mu}{\sigma}.
    \label{eq:zscore}
\end{equation}

\noindent This equation defines the $\sigma$-variance of a point $x$ from the mean of the population $\mu$. This value can be calculated for a data point (halo) or can be generalized to a sample mean (candidate subpopulation). The Mahalanobis distance, which extends the \rm{Z}-score to multiple variables, can then be described by the covariance matrix $S$, multivariate mean $\bm{\mu}$, and sample points $\bf{x}$. The distance is defined as \citep[see, e.g.,][]{DeMaesschalck2000,GeunKim2000}

\begin{equation}
    d_{\rm{M}} = \sqrt{(\mathbf{x}-\bm{\mu})^{T} S^{-1} (\mathbf{x}-\bm{\mu})}.
    \label{eq:mahal}
\end{equation}

\noindent Once again this value can be extended from a point (halo) to the mean of a sample. The Mahalanobis distance has previously been used in an astrophysical context in \citet{Blaylock-Squibbs2023}. In this paper the matrix representation of mean, $\bm{\mu}$, is defined as the halo population means for each feature presented in Section~\ref{subsec:var}. We define the sample population of halos as the DCBH candidates. Mahalanobis distances with a value $>1$ are generally considered to show a low similarity between the population and sample distributions. To minimize the effect of outliers on data we work in log-space because the range of fields often exceed two orders of magnitude.

With the Mahalanobis distance we perform recursive feature selection. Starting with all the features present we can test the distance as each variable is removed and eliminate the variable that causes the least change in distance. When the removal of a feature causes minimal change to the distance it likely does not affect determining the candidacy of a halo. The process can be performed until only the most important features remain. This recursive selection can also be implemented in reverse order by eliminating the most important (largest changes in distance) first; we found this decreases the effects of correlated variables which is expanded upon in Section~\ref{subsubsec:corrmeth}.

\subsubsection{Recursive Feature Ranking} \label{subsubsec:rec}
With \texttt{sklearn} we perform recursive feature ranking, a supervised machine learning method. We use the Logistic Regression classification method that evaluates the probability of an outcome in a binary classification system, candidate or non-candidate, by using an logistic probability function \citep{Guyon2002,Musa2013}. This method produces a feature ranking based on importance in the probability function. An additional method uses decision trees \citep{Breiman2001,Chen2020}, which attempts to separate the populations of candidate and non-candidate halos by recursively dividing the sample space.

\subsubsection{Permutation} \label{subsubsec:perm}
In addition to recursive methods, \texttt{SciKit-Learn} offers permutative feature ranking. Permutation-based ranking assesses features based on how a selected model's performance changes when a feature is removed \citep{Zien2009,Konig2021}. The more the model performance decreases when a feature is removed, the greater its importance in determining the class (candidacy) of the data point (halo). We use a Random Forest Classifier model that applies decision trees to classify a target point. We chose this classification method as we expect important features to have distinct separations between candidates and non-candidates. However, we do not make use of any robust analysis to determine the best model; instead we leave this to future work. 

\subsubsection{Correlated Features} \label{subsubsec:corrmeth}
Many of the features are expected to be correlated. Some variables, such as density, are contained within the physical definition of other features. Others have predicted relations, such as that of LW flux and distance to closest galaxy. Thus, dependencies must be included in our analysis for a more thorough understanding of feature importance.

To reduce the effect of correlated variables on our analysis, we use the \texttt{hierarchy.ward} function provided by \texttt{sklearn} to sort features into groups based on their distance, where the distance is defined as $d = 1-\text{Cor}(X,Y)$ between two variables $X$ and $Y$. Using these groups we can perform adjusted Mahalanobis and permutation methods. 

Whereas the \rm{Z}-score evaluates the variation of each variable separately, the Mahalanobis distance relies on the covariance matrix; thus features with multiple strong correlations are weighted higher. To revise the Mahalanobis ranking method provided in Section~\ref{subsub:mahalmethod} to accommodate correlations between features, we can evaluate and eliminate in groupings.

To perform permutation ranking on groups we use the function provided by \citet{Plagwitz2022} as a modification of \texttt{SciKit}'s permutation function to evaluate group importance while also taking individual feature importances into account. 

\section{Results} \label{sec:res}
\subsection{Probability Distributions} \label{subsec:prob}
\begin{figure*}[htb!]
    \centering
    \includegraphics[width=\textwidth]{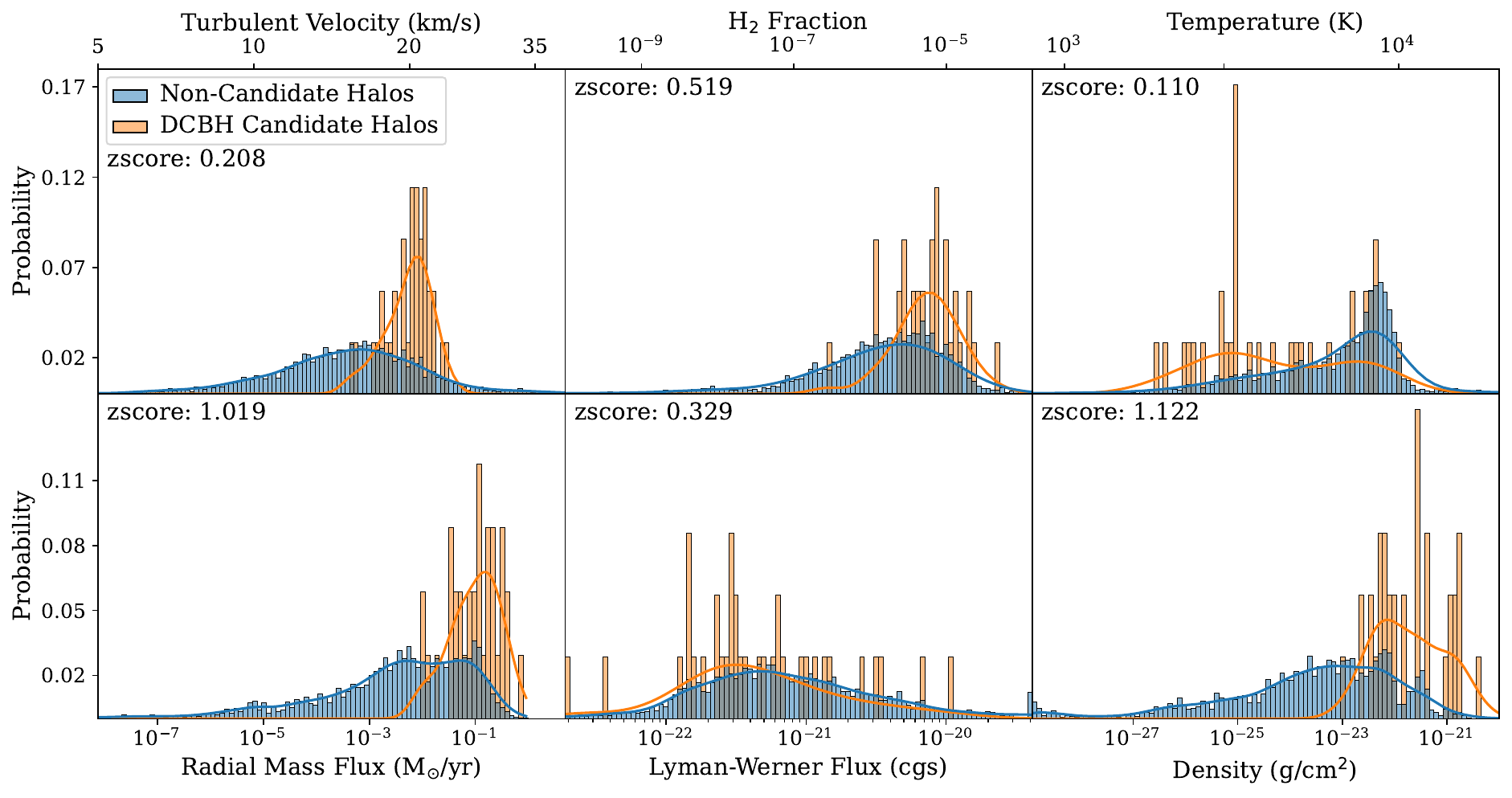}
    \includegraphics[width=\textwidth]{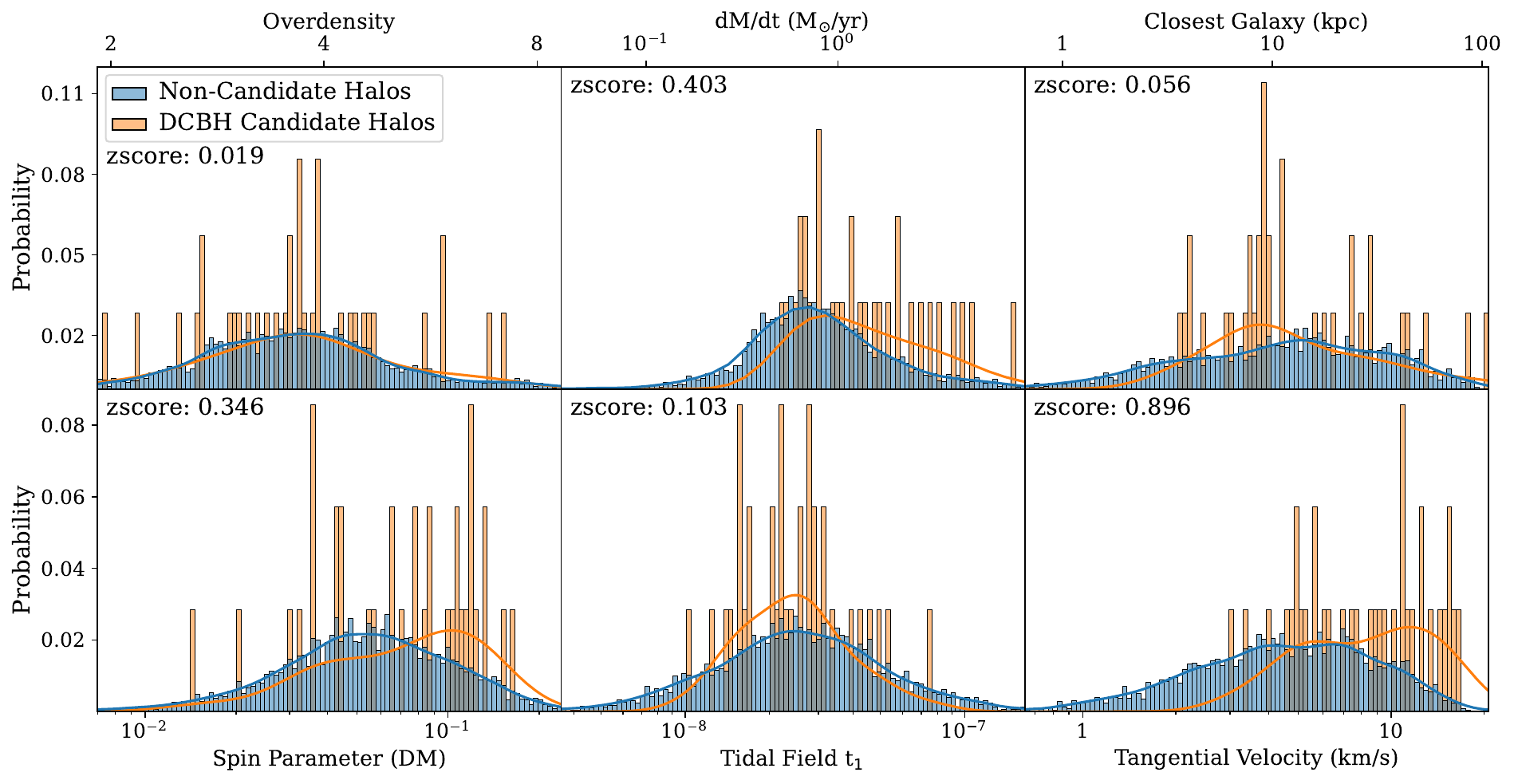}
    \caption{Probability distribution comparisons for candidate (orange) vs. non-candidate (blue) halo populations. The histograms display binned probabilities and kernel density estimate functions. Individual feature \rm{Z}-scores are provided in the upper left-hand corners of each plot. The plots show regions of peak probability with outliers removed; in the case of most plots between 20-50 (non-candidate only) halos are removed where the probability distribution decreases to $<0.005$ in outlying regions. The $\rm{H}_{2}$ fraction and temperature distributions have removed 300 and 500 halos, respectively, as the distribution functions have long tail ends. This removal serves to better compare the distributions. Some features (average growth over redshift, gas spin parameter, and radial velocity) are excluded as all are highly correlated with another variable so their inclusion is unnecessary. The cgs units for LW flux are $\rm{{erg} \, {cm}^{-2} \, s^{-1} \, \rm{Hz}^{-1}}$}.
    \label{fig:hists}
\end{figure*}

Figure~\ref{fig:hists} displays probability distributions comparing candidate versus non-candidate halos along with the individual feature \rm{Z}-scores. At a glance we can see that some candidate distributions are noticeably offset from that of the non-candidate population. The density is on average greater for candidates, with an approximately $1.1\sigma$ deviation of the mean candidate halos from the mean halo population, which is expected as these halos are collapsing as they cross the atomic cooling limit. Additionally, the mean candidate radial mass influx mean has a pronounced difference of $\sim\! 1\sigma$ showing a higher influx of mass in candidates. Furthermore, tangential velocity seems to be larger for candidate halos, suggesting an importance on halos with more rotation. While having only a mid-range \rm{Z}-score, turbulent velocity for candidates is strongly peaked at $20 \, \rm{{km} \, s^{-1}}$ and the non-candidate distribution has a much broader range.

Many features do not exhibit a deviation of the candidate sample from the general population distributions, indicating lesser feature importance. The LW Flux, originally thought necessary in dissociating $\rm{H}_2$ to prevent cooling and fragmentation, has in recent years been superseded by other dissociation mechanisms \citep{Fernandez2014,Patrick2023}. From our figure we see that the candidate and non-candidate distributions for LW flux are relatively similar in shape and the \rm{Z}-score of $0.329$ represents a comparatively lower candidate mean to population mean. This corroborates the idea that LW flux is not prominent in candidate halos. The main source of LW flux in our halo sample is from stars in other galaxies, thus we expect a close relation between LW flux and distance to nearest galaxy. The very small \rm{Z}-score of distance agrees with the low importance of LW flux and disagrees with the existence a "Goldilocks zone" for DCBH formation. This zone had also been a popular theory until recently in which halos were close enough to a galaxy to have a strong source of LW radiation but not to be enriched by the galaxy \citep{Regan2017}. These two fields are also expected to be highly correlated to $t_{1}$ due to their dependence on outside sources. The tidal field $t_{1}$ of the candidates has an $\sim\! 0.1\sigma$ deviation from the population, furthering the evidence against the importance of surrounding galaxies on DCBH hosts. Moreover, the overdensity has a particularly low \rm{Z}-score, which contributes to the overall idea of insignificant halo environment effects on candidacy. While overdensity is not important in differentiating halos, all of the halos -- candidates and non-candidates alike -- form in overdense regions, an aspect of the Rarepeak region of the simulation domain. 

\subsection{Feature Selection} \label{subsec:resfeat}
\subsubsection{Recursive Ranking and Decision Tree} \label{subsubsec:resrec}
The performance of the decision tree behaved as expected, first making cuts on metallicity and halo mass -- two variables used in the selection process for DCBH candidates. Other important variables in the decision tree included: temperature, growth rate, turbulent velocity, tangential velocity, and LW flux. As the depth increases, it becomes more difficult to differentiate candidates from non-candidates; so more small divisions are made and features used to make these later cuts are likely not as important. 

Using recursive Logistic Regression classification we generate a feature importance ranking. The results of which are contained in the ranked list below: 
\begin{enumerate}
    \bfseries
    \itemsep0em
    \item Metallicity
    \item Halo Mass
    \item Radial Mass Flux
    \item Tangential Velocity
    \item $\mathbf{dM/dt}$ (Halo Mass Growth)
    \item Density
    \item Overdensity
    \item $\mathbf{dM/dz}$ (Halo Mass Growth)
    \item $\mathbf{\rm{H}_{2}}$ Fraction
    \item Tidal Field $\mathbf{t_{1}}$
    \item Turbulent Velocity
    \item Radial Velocity
    \item Distance to Nearest Galaxy
    \item Stellar Mass
    \item Lyman Werner Flux
    \item Spin Parameter (Gas)
    \item Spin Parameter (DM)
    \item Temperature
\end{enumerate}

As expected from our candidacy selection method, metallicity and halo mass are again high importance features along with radial mass flux, tangential velocity, growth rate, and density. These features suggest that candidates tend to be more massive with high growth influxes. Features of minimal importance include temperature, spin parameters, and LW flux, agreeing with the recent literature \citep[e.g.,][]{Regan2020a,Regan2020c} as well as the previous tests. Also of low importance is stellar mass which does not match with the candidacy selection method. However, this follows from the cut made on the total halo population to include only halos with no young (and thus) massive stars. This limits the effects of interior halo stars since it is thought that direct collapse black holes form in halos with minimal star activity to avoid cooling and fragmentation.

\subsubsection{Correlated Variables} \label{subsubsec:corr}

\begin{figure*}[htb!]
    \includegraphics[width=\textwidth]{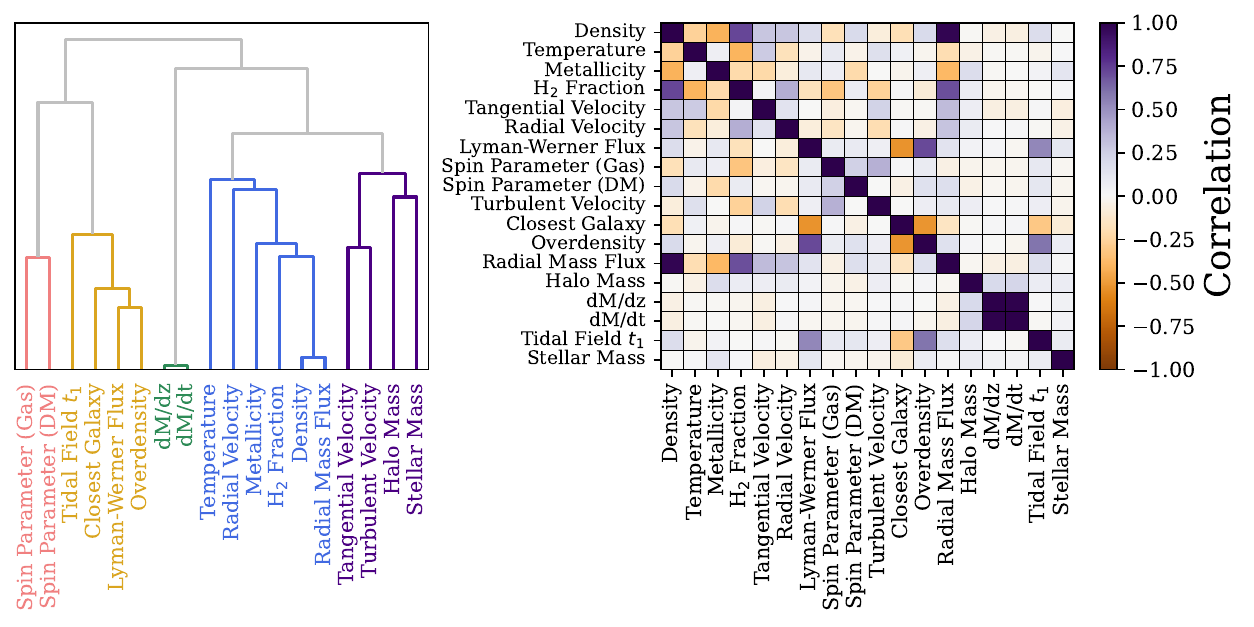}
    \caption{The left panel displays a hierarchy linkage map between features based on the Spearman correlation coefficient distance \citep[e.g.,][]{Yu2024}. The groups are numbered from left to right in the diagram and designated by colors with Group 1 in pink, Group 2 in yellow, Group 3 in green, Group 4 in blue, and Group 5 in purple; these numbers and colors will be used throughout the rest of this paper. The right panel shows the correlation matrix between all variables used in feature selection, where $1$ is a high positive correlation, $-1$ is a high negative correlation, and $0$ is no correlation.}
    \label{fig:corr}
\end{figure*}

The correlation matrix in the right panel of Figure~\ref{fig:corr} displays how the features of interest are correlated. The linkage diagram in the left panel of Figure~\ref{fig:corr} displays the process of how the features break into groups based on the Spearman correlation coefficient between features. The groups are as follows:
\begin{enumerate}
    \itemsep0em
    \item \textbf{Group 1}: Gas Spin Parameter and Dark Matter Spin Parameter
    \item \textbf{Group 2}: Tidal Field $t_1$, Distance to closest galaxy, LW Flux, and Overdensity
    \item \textbf{Group 3}: Growth Rates (dM/dt and dM/dz)
    \item \textbf{Group 4}: Temperature, Radial Velocity, Metallicity, $\rm{H}_{2}$ Fraction, Density, and Radial Mass Flux
    \item \textbf{Group 5}: Tangential Velocity, Turbulent Velocity, Halo Mass, and Stellar Mass
\end{enumerate}

Groups (1) and (3) each describe values that are similar or approximately the same. Group (1) contains the dark matter and gas spin parameters suggesting that one value is sufficient in understanding spin differences in candidates and non-candidates. Growth rates with respect to time and redshift comprise group (3); these features are at a correlation of $\sim\! 1$, which is expected as they measure the same property with only slight variations due to smoothing and the conversion factor between time and redshift. Group (2) involves features dependent on the halo environment. The overdensity describes how dense is the surrounding $15 \, \rm{kpc}$ radius region and the distance to the closest galaxy measures the closeness of the halo to other strong radiation sources. These two features both quantify if the halo's environment is sparse or populated with galaxies. The other two features in this group are the LW flux and the tidal field $t_{1}$, both dependent on the characteristics of the halo's surroundings. In more dense regions where the halo is close to another galaxy, the LW flux is expected to increase from stars in nearby galaxies and $t_{1}$ would increase from the tidal forces exerted by the associated large-scale structure. In the correlation diagram of Figure~\ref{fig:corr} we see tidal field $t_{1}$, LW-flux, and overdensity are all strongly positively correlated and all three are negatively correlated to closest galaxy as large distances to galaxies means less stellar radiation, less tidal force, and a less dense region. Group (4) describes properties of the central halo. Density shows a strong correlation to many features due to its importance in halo evolution and the dependency of other features on it (e.g., see Equation~(\ref{eq:rmf})). Group (5) contains halo characteristics. In the upcoming sections we perform feature selection based on the groups presented in the left panel of Figure~\ref{fig:corr}.

\subsubsection{Mahalanobis Distance} \label{subsubsec:resmahal}

\begin{figure*}[htb!]
    \centering
    \includegraphics[width=\textwidth]{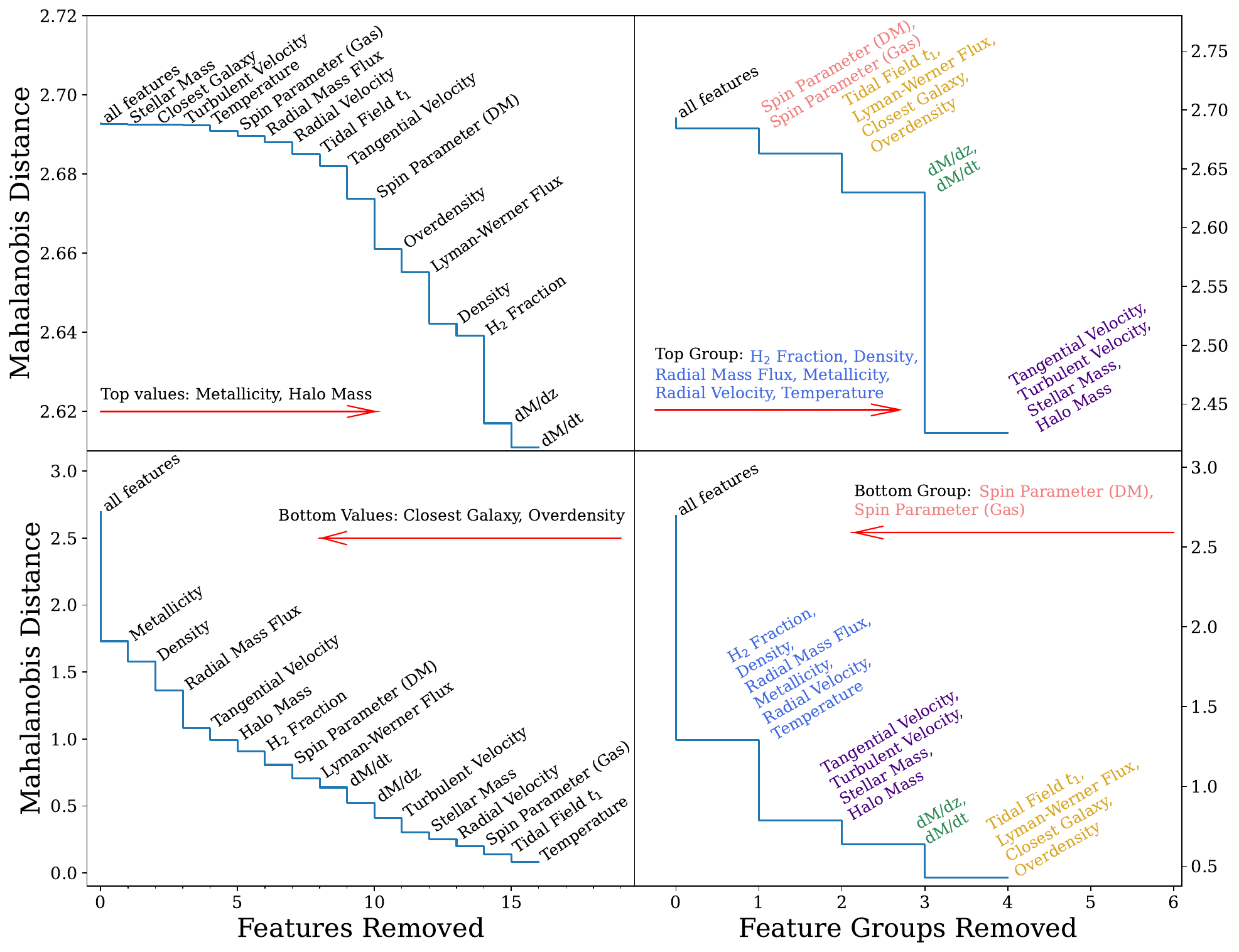}
    \caption{Stair plots displaying the effect on the Mahalanobis score as features are progressively removed. The top row shows recursive elimination of least important features and contain the most important values in the lower left corners. The bottom row shows recursive elimination of the most important features with the least important labeled in the top right. In the left column features are removed individually, and in the right column, features are removed as correlated groups (information on how groups are made in Section~\ref{subsubsec:corr}). The red arrow designates the direction of importance.}
    \label{fig:mahal}
\end{figure*}

The stair plots shown in the left panels of Figure~\ref{fig:mahal} display how the Mahalanobis distance changes as features are recursively removed individually. Both removal directions agree on metallicity as a high importance feature, following from the selection method for candidate halos. Other features that both methods select as important are halo mass and density. However, the rankings are not in complete agreement. For example, radial mass flux is selected as high importance when features are removed from most to least important but is toward the lower end of importance when features are removed from least to most important. Similar trends can be seen with growth rate variables. This is likely due the dependence of the Mahalanobis distance on the covariance matrix. Features strongly correlated to others such as LW radiation can be falsely conflated as important (see Section~\ref{subsubsec:corr}). Specifically, it seems this is increased by high correlations to unimportant variables.

The right panels of Figure~\ref{fig:mahal} present the Mahalanobis ranking test performed on feature groups, chosen by the process outlined in Section~\ref{subsubsec:corr}. When implemented using groups the elimination process is the same for removal in both directions. Both processes select the group containing density, metallicity, and radial mass flux among other features as the most important which agrees with our candidacy selection and expectations. The least important group is the spin parameter group suggesting its minimal role in DCBH formation.

However, some information is lost when grouping the features. These groups contain up to six features that can vary in importance; using this method, unimportant features can be given a higher ranking or vice versa. Thus, these results should be weighed in conjunction with other statistical and machine learning methods.

\subsubsection{Permutation Ranking} \label{subsubsec:resperm}

\begin{figure*}[htb!]
    \centering
    \includegraphics[width=\textwidth]{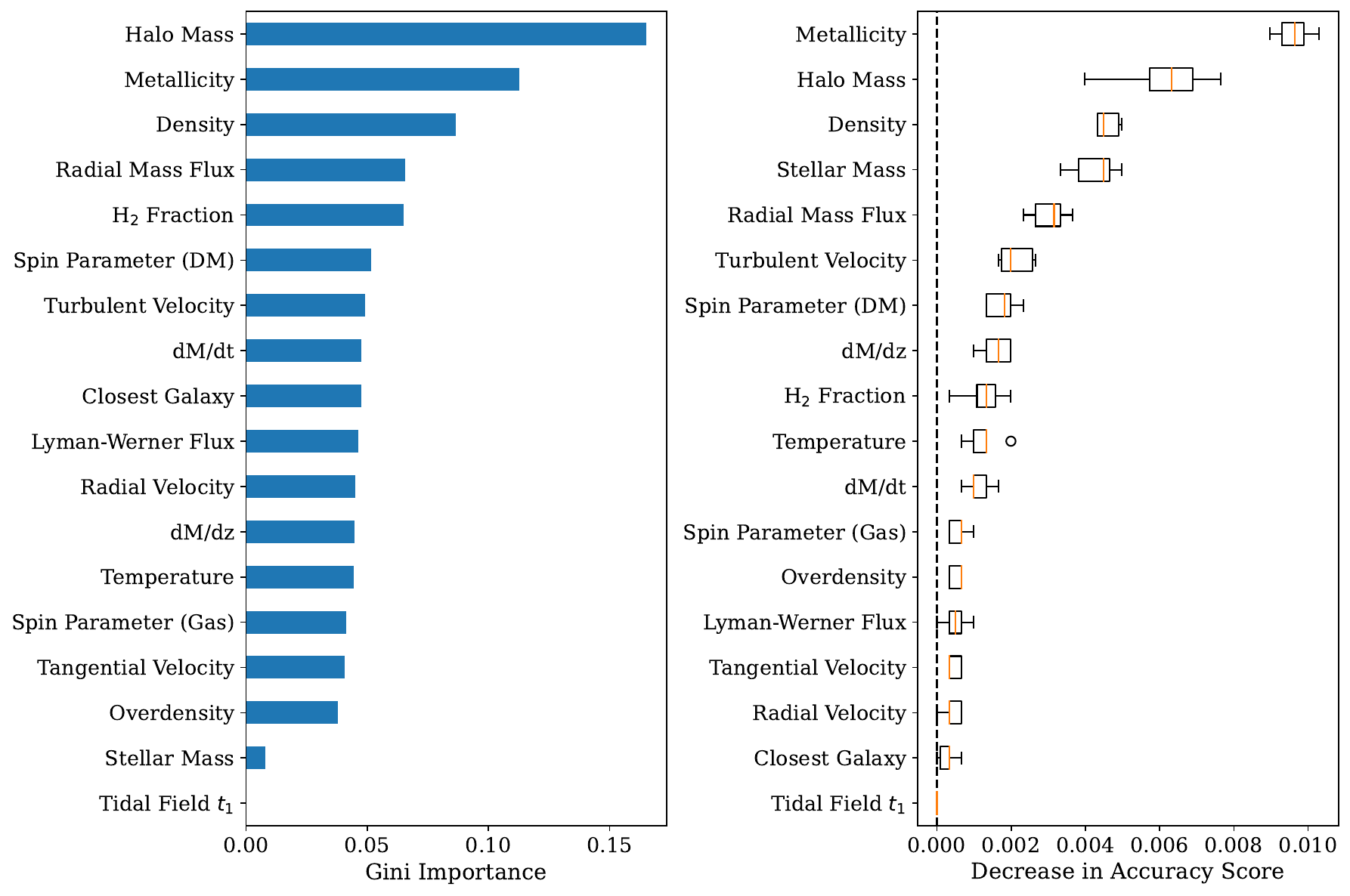}
    \caption{The left panel displays the Gini importances \citep{Breiman1984,Nembrini2018} of each feature using a random forest classifier model with a higher score indicating more importance. The right panel shows box plots for the decrease in accuracy score of the model when each feature is progressively removed, with larger decreases indicating more importance.}
    \label{fig:permutation}
\end{figure*}

Figure~\ref{fig:permutation} displays the Gini importance of each feature (left) along with the decrease in accuracy score when the corresponding feature is removed from the model (right). The model trained on the data is a Random Forest Classifier. Features that have a high Gini Importance or yield a large decrease in the accuracy score when removed are deemed important. From Figure~\ref{fig:permutation} we can see that both rankings using Gini Importance and decrease in accuracy select halo mass, density, and metallicity as top-ranked features. This result agrees with the candidate selection method as well as other ranking processes in previous sections. Similarly, both rank the tidal field, spin parameters, and the distance to the closest galaxy low. However, they disagree on the rankings of the features in between. This is not unexpected due to the large number of features and the large range of values that each feature can take on. Owing to this as well as the correlations between features, it is difficult to distinguish rankings with little difference between variables; this is where grouping the features can help. 

\begin{figure}[htb!]
    \includegraphics[width=0.45\textwidth]{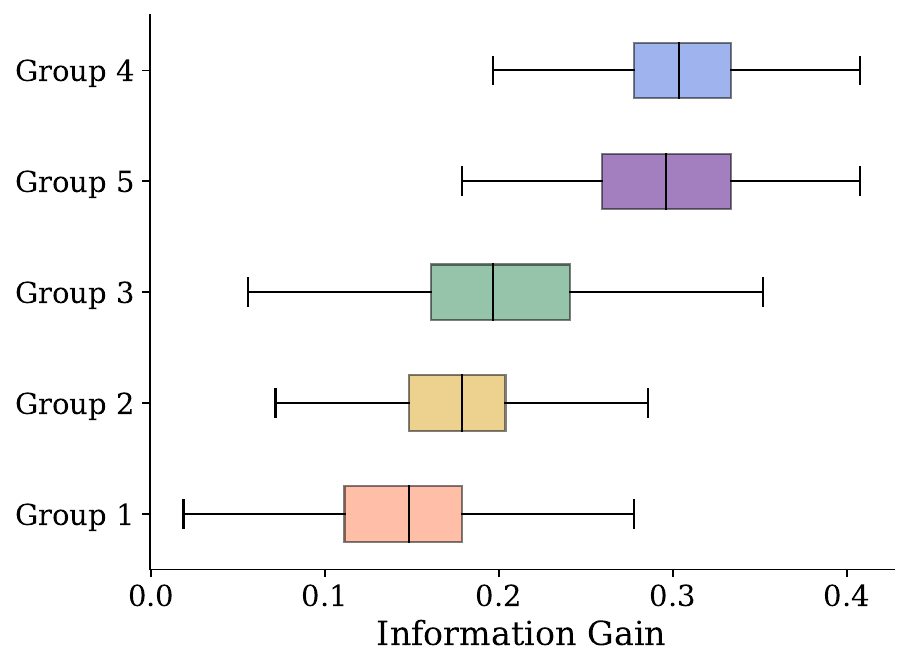}
    \caption{Information gain produced by each feature group where the groups are as follows: (1) Gas and Dark Matter Spin Parameter, (2) Tidal Field $t_{1}$, Distance to closest galaxy, LW Flux, and Overdensity, (3) Growth Rates (dM/dt and dM/dz), (4) Temperature, Radial Velocity, Metallicity, $\rm{H}_{2}$ Fraction, Density, and Radial Mass Flux, and (5) Tangential Velocity, Turbulent Velocity, Halo Mass, and Stellar Mass.}
    \label{fig:grpperm}
\end{figure}

Figure~\ref{fig:grpperm} shows the information gain that each group (grouping process and outcomes are described in Section~\ref{subsubsec:corr}) yields in a machine learning model. The method followed to obtain these values recursively permutes between feature groups and their constituents for the most inclusive results. From this figure we have the ranking of high to low importance as Groups (4), (5), (3), (2), (1); this ranking order of the groups agrees directly with the results of the grouped Mahalanobis test results in Figure~\ref{fig:mahal}. We can distinguish two regions: high importance -- groups (4) and (5) -- and lesser importance -- groups (3), (2), and (1). The bottom two groups represent environmental factors and spin parameters, respectively. Group (3), the group of middle importance, contains the halo growth rates. The two groups of highest importance include our candidacy selection variables in addition to prominent features of the halo's central region, specifically those pertaining to mass and halo dynamics; this suggests DCBH formation is dependent on fast, substantial core growth and strong motions in the central halo. Due to the large number of halos within our data set as well as a large range of values for each feature group the test produces large error ranges, shown by the $90\%$ confidence interval around the median information gain. Thus, the distinctions between feature groups within the top or bottom regions are not absolute and the valuable information to take from this figure is that groups (4) and (5) are the most important in determining candidacy. It is important to note these two groups contain $\sim\! 50\%$ of the halo features, and from this we cannot distinguish exact feature importances; the ranking may be driven by one variable and the others may belong at a different ranking. Therefore, this does not mean that all features contained in Groups (3), (2), and (1) are necessarily unimportant.

\subsection{Halo Property Comparison} \label{subsec:halocompare}
In this section we present a more physical view into differences between candidate and non-candidate halos. First, we consider all halos in our data set and then we select a case study of two halos in the upper mass range of our sample -- one non-candidate and one candidate -- close to each other in LW flux/mass phase space; chosen feature values are presented in Table~\ref{tab:DCBHvCan}. These two features are chosen as they have both been used in the past to determine potential candidacy, with halo mass as one of our requirements.

\begin{table}[htb!]
    \caption{\label{tab:DCBHvCan}}
    \vspace{-5mm}
    \begin{center}
        \par{Halo case study features}
    \end{center}
    \vspace{-5mm}
    \begin{ruledtabular}
    \begin{tabular}{lcr}
      Feature & DCBH Candidate & Non-Candidate \\
    \colrule
    Halo Mass $(\rm{M}_{\odot})$ & $6.28 \times 10^{7}$ & $6.19 \times 10^{7}$ \\
    LW Flux (cgs) & $2.42 \times 10^{-21}$ & $2.35 \times 10^{-21}$ \\
    Density $(\rm{g \, {cm}^{-3}})$ & $1.69 \times 10^{-21}$ & $1.91 \times 10^{-25}$ \\
    Temperature (K) & $4230$ & $9040$ \\
    $\rm{H}_{2}$ Fraction & $2.84 \times 10^{-6}$ & $5.86 \times 10^{-7}$ \\
    RMF $(\rm{M}_{\odot} \, \rm{{yr}^{-1}})$ & $0.2345$ & $0.0002$
    \end{tabular}
    \end{ruledtabular}
    \tablecomments{RMF designates the radial mass influx. Values are as calculated in Section~\ref{subsec:var}.}
\end{table}

\begin{figure}[htb!]
    \includegraphics[width=0.45\textwidth]{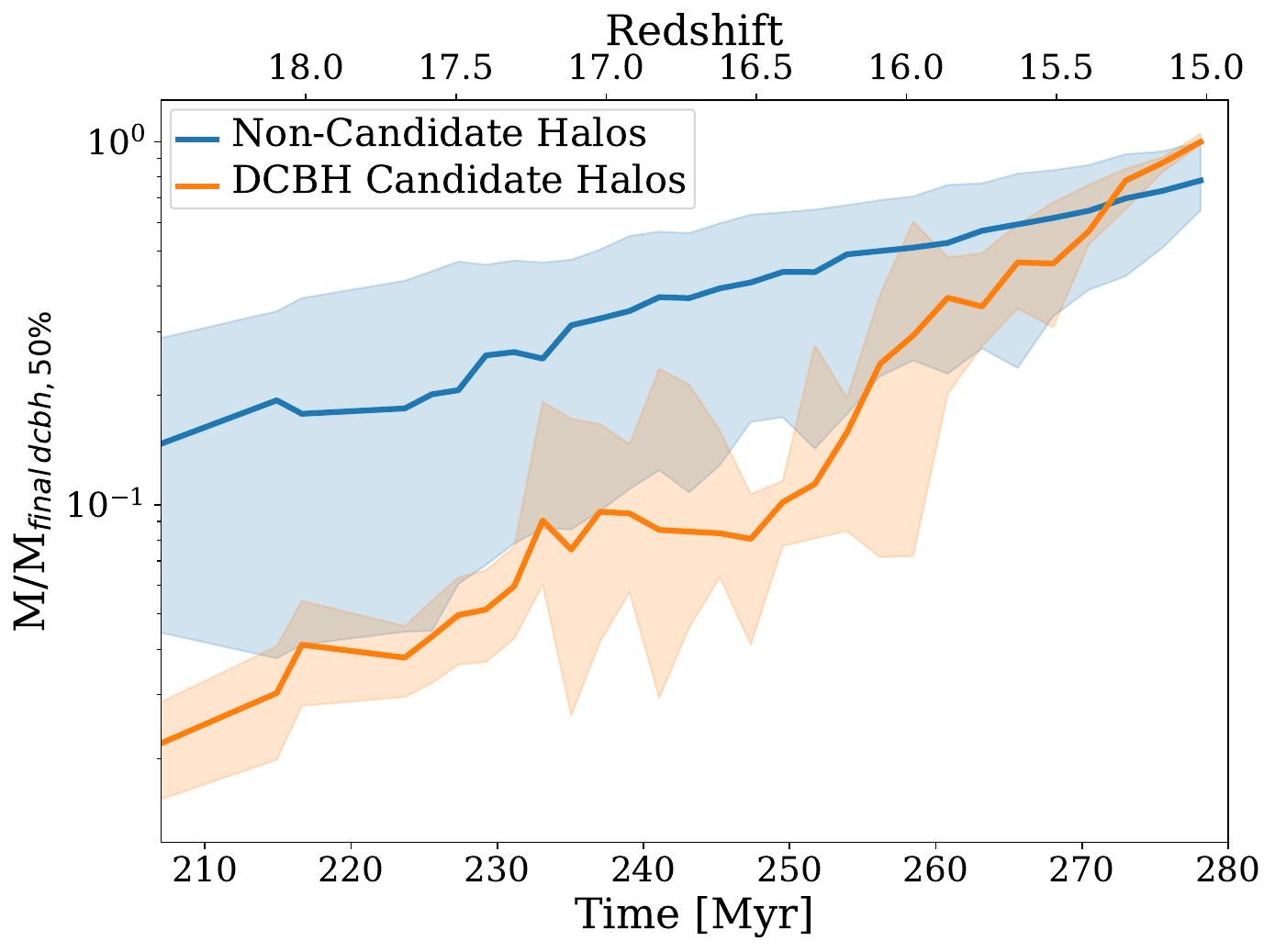}
    \caption{Comparison of normalized halo mass with respect to time for non-candidate halos (blue) and candidate halos (orange) with a final redshift of $z = 15.0$. Both samples are normalized to the final median DCBH mass. The lines represent the median values of each population with the shaded regions showing the interquartile ranges.}
    \label{fig:growthratecomp}
\end{figure}

Figure~\ref{fig:growthratecomp} displays the entire subset of halos at redshift $z = 15$ and compares non-candidate versus candidate growth. The lines show the median mass of each class at each point in time normalized by the final median mass of the candidate halos. From this figure we can see that overall the non-candidate (blue) growth is much flatter, starting at a higher original mass but gaining less mass over time than the candidates (orange); the mass growth of the candidates starts off with a slope comparative to that of the non-candidates but begins to more rapidly grow at $\sim\! 247 \, \rm{Myr}$ (redshift of $z \sim\! 16.5$). From this time of increased candidate growth to the final output time the non-candidates have a $\sim\! 1.9 \times$ mass increase whereas the candidates grow by $\sim\! 12.4 \times$, more than six times the growth of non-candidates. Thus, over $\sim\! 130 \, \rm{Myr}$ the non-candidates have only doubled in size when the candidates grow by 12 times their original mass. The non-smoothness of the candidate halo growth is both due to the characteristics of fast growing halos and a lower number of halos in the candidate subset which leads to more spread and less smoothing from averaging. 

\begin{figure*}[htb!]
    \centering
    \includegraphics[width=\textwidth]{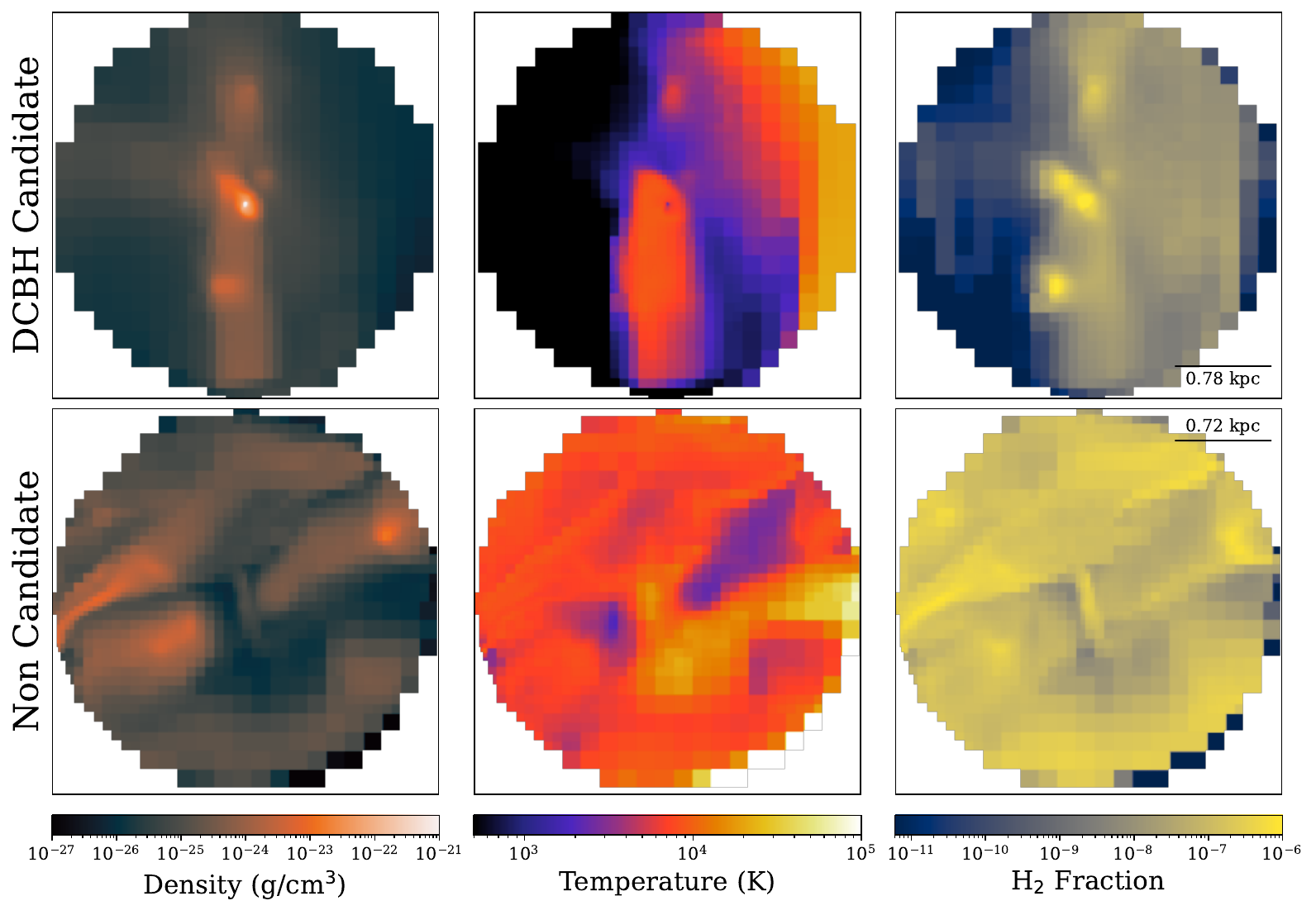}
    \caption{A visual comparison of a DCBH candidate halo (top, $M_{\rm{vir}} = 6.28 \times 10^{7} \, \Msun$, LW flux $= 2.42 \times 10^{-21} \, \rm{cgs}$) and a non-candidate halo (bottom, $M_{\rm{vir}} = 6.19 \times 10^{7} \, \Msun$, LW flux $= 2.35 \times 10^{-21} \, \rm{cgs}$). Slices are through the halo center for three main halo variables: density (left), temperature (middle), and $\rm{H}_{2}$ fraction (right). Scale bar in right-most plots for each halo represents the halo's virial radius and the field of view is 4 times the virial radius.}
 \label{fig:halos}
\end{figure*}

Moving onto the case study, Figure~\ref{fig:halos} displays slices of density, temperature, and $\rm{H}_{2}$ fraction for a non-candidate halo (bottom) and a candidate halo (top). These halos are chosen to be close in LW flux/mass phase space with a high mass. From the density slices we can see that the non-candidate halo has lost much of its gas to filamentary structures outside the central halo whereas the candidate halo has a dense central region and minimal loss to filaments. When exploring the star formation history of the halos we determine that the candidate contains no stars (a requirement for candidacy) whereas the non-candidate halo has experienced star formation starting $\sim\! 41 \, \rm{Myr}$ before this snapshot. Additionally, as shown in the temperature slices, the non-candidate halo has heated the surrounding region likely by stars, whereas the core of the candidate halo is heated through virialization and compression. Furthermore, the non-candidate has relatively higher $\rm{H}_{2}$ fraction in diffuse gas, indicative of previously being photoionized, i.e. a relic \ion{H}{2} region. On the other hand, only the candidate halo's core has a boosted $\rm{H}_{2}$ fraction more typical of a primordial halo collapsing for the first time. That being said, the $\rm{H}_{2}$ fraction in both halos is small enough so that the cooling time is longer than the free-fall time, suggesting that $\rm{H}_{2}$ fraction is unimportant in candidacy. 

\begin{figure*}[htb!]
    \centering
    \includegraphics[width=\textwidth]{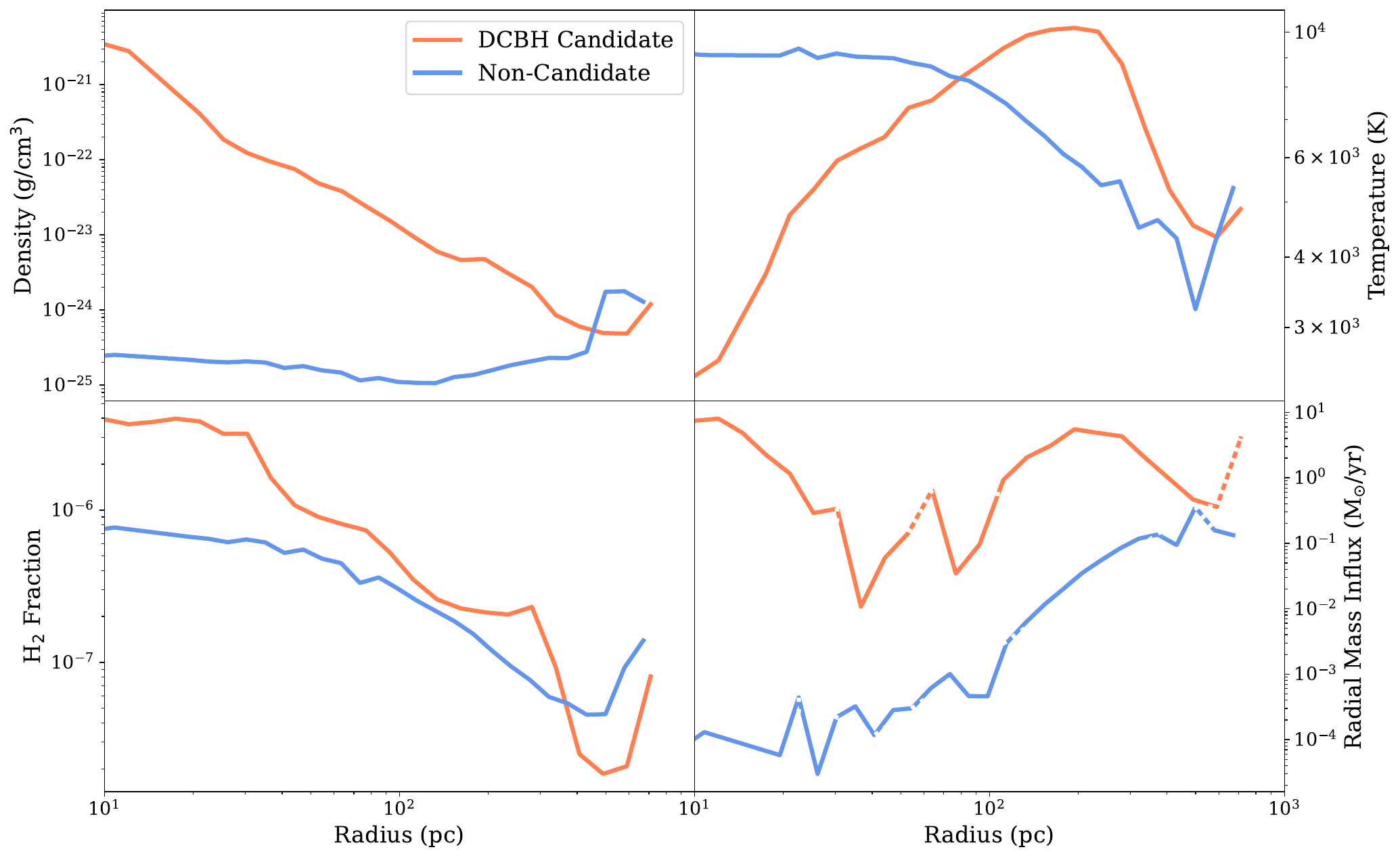}
    \caption{Radial profiles for density (top left), temperature (top right), $\rm{H}_{2}$ fraction (bottom left), and radial mass flux (bottom right) for a DCBH candidate (orange) and non-candidate (blue). Dashed lines in the radial mass flux profile represent radii of mass outflux.}
 \label{fig:radprofs}
\end{figure*}

Figure~\ref{fig:radprofs} shows radial profiles for the same candidate (orange) and non-candidate halo (blue) as the previous figure in density, temperature, $\rm{H}_{2}$ fraction, and radial gas mass flux. For the features represented in Figure~\ref{fig:radprofs} the corresponding central averages can be found in Table~\ref{tab:DCBHvCan}.  The differences between the two cases are clear in the profiles, as stellar feedback from the prior star formation event in the non-candidate halo evacuated its dense core.  

The candidate halo has an isothermal profile ($\rho \propto r^{-2}$) in its envelope between the $100 \, \rm{pc}$ core and the virial radius ($780 \, \rm{pc}$).  Within the core, the $\rm{H}_{2}$ fraction increases by an order of magnitude to $4 \times 10^{-6}$ in the center, gradually cooling the gas from $10^{4} \, \rm{K}$ in the envelope to $2000 \, \rm{K}$ at the center.  Within the innermost $20 \, \rm{pc}$, the temperature profile is steeper, causing the density profile to steepen as it loses pressure support.  The halo core is collapsing with an average gas influx of $0.1 \Msun \, \mathrm{{yr}^{-1}}$ between $30-100 \, \rm{pc}$ and rising to $5 \Msun \, \mathrm{{yr}^{-1}}$ in the center.

In stark contrast, the non-candidate halo has a nearly uniform density profile with a central temperature of $8000 \, \rm{K}$, dropping to $5000 \, \rm{K}$ at the virial radius, in accordance with a relic \ion{H}{2} region and supernova remnant.  The halo center has just started to recover from the feedback episode with a slight gas influx, fluctuating in the range $10^{-3}-10^{-4} \Msun \, \mathrm{{yr}^{-1}}$, a few orders of magnitude lower than the candidate halo.


\section{Discussion} \label{sec:dis}

JWST observations of $z \gtrsim 6$ AGN \citep{Inayoshi2020,Goulding2023,Larson2023,Bogdan2024,Maiolino2024c} place strong constraints on SMBH early growth but less so on their formation mechanisms. Regardless of whether their seeds are light or heavy, rapid growth must occur at a high duty cycle, requiring a near continuous and ample gas supply to the (proto-)galactic center. We have demonstrated that the central halo core quantities are the best indicators for an ensuing collapse. In this work, we considered these sites to be optimal for DCBH formation; however, this collapsing gas-rich core can also form a dense stellar cluster or supply accretion flows onto light seeds.

We note that the observational research research cited above reaches very similar conclusions about black hole growth despite observing different populations of black holes. The papers \citet{Goulding2023}, \citet{Larson2023}, and \citet{Bogdan2024} discuss black holes of $10^{7} \, \rm{M}_{\odot} \lesssim M_{\bullet} \lesssim 10^{8} \, \rm{M}_{\odot}$, though \citet{Maiolino2024c} presents observations of a population of black holes covering a much wider range from $10^{5.5} \, \rm{M}_{\odot} \lesssim M_{\bullet} \lesssim 10^{8} \, \rm{M}_{\odot}$. The former provide much stronger evidence for the DCBH formation pathway (or super-Eddington accretion for lighter seeds). The results from \citet{Maiolino2024c} give more leeway to a broader distribution of (lower) seed black hole masses, as is also discussed in the review article by \citet{Inayoshi2020}, and though they note that many of the black holes in their sample have low Eddington ratios ($\lambda_{\rm{Edd}} \lesssim 0.5$), the lower mass black holes of $M_{\bullet} \lesssim 10^{6} \, \rm{M}_{\odot}$ tend to accrete much closer to the Eddington limit. Therefore, even though these papers look at different populations of AGNs, the conclusions reached by both sets of papers are consistent: the observed population of black holes can be explained through a combination of light seeds, which would be required to accrete at much higher (typically super-Eddington) rates, and heavy seeds, which may be allowed to accrete at lower (likely sub-Eddington) rates. These findings cannot distinguish between the two seeding scenarios, but the presence of $M_{\bullet} \sim 10^{8} \, \rm{M}_{\odot}$ SMBHs in the early universe does constrain that black hole accretion rates must have been near or above the Eddington limit, at least for some seed black holes.

The commonality between the two seeding scenarios is the dependence on gas supply. The light seeds must undergo epochs of super-Eddington accretion events and remain centered in the galaxy where dense gas is most likely to exist \citep{Lupi2024a,Lupi2024b,Mehta2024}. An intense halo growth history with enough gas-rich mergers may supply sufficient fuel for the light seed to grow to $M_{\bullet} \gtrsim 10^{5} \Msun$. DCBHs, and heavy seeds in general, forego the difficulty of initial growth up to these scales, but theoretical and computational work have shown that their formation could be rare \citep{Habouzit2016,Regan2020a,Volonteri2021,McCaffrey2024}.  Our work has highlighted a rapid halo growth history as an important feature in the candidate halos that could induce DCBH formation or further feed the light seeds, given that the progenitor halos are gas-rich.

Halos that are prime candidates for DCBH and/or dense stellar cluster formation require star formation to be suppressed through some mechanism when their progenitors are below the atomic cooling limit. As shown in Section \ref{subsec:halocompare} and many previous studies \citep[e.g.][]{Wise2012,Kimm2017,Katz2020,Trebitsch2021,Brauer2024,Kiyuna2024}, stellar feedback has a dramatic effect on the cold dense gas reservoir in these fragile cradles. Once a halo reaches the atomic cooling limit or thereabouts without star formation, its $\sim\!10^{5} \Msun$ core can cool and collapse. Whether it forms a dense stellar cluster or a single supermassive star and a subsequent DCBH depends on the thermodynamics of the collapse and ensuing fragmentation, if any. Recent simulations have shown the resulting object is an amalgamation, being a stellar cluster with several stars having masses greater than $1000 \Msun$ \citep{Wise2019,Regan2020c}. The subsequent evolution of such a cluster is an ongoing topic of research, but such an environment is prime for heavy seeding and possible BH mergers. A recent model presented by \citet{Kroupa2020} suggests a formation mechanism in which the dense stellar cluster leaves a black hole subcluster behind after stellar death which proceeds to collapse into a single SMBH seed.

\subsection{Applicability and Implications}

Our work can be applied to both the light or heavy seeding scenarios. Our analysis identifies halos that are susceptible to strong central inflows. Therefore, these halos may host a massive black hole once they reach the atomic cooling limit, either through a direct collapse or rapid accretion. The primary motivation of this work is to identify the most important features in DCBH formation to create a more physically motivated seeding model for semi-analytic galaxy formation models or subgrid models in cosmological simulations. The results in this paper only represent the first step by identifying the features. We are currently finalizing a probabilistic formation model based on this work that identify the criteria, i.e. decision boundaries in a support vector machine, to trigger DCBH formation (Pries et al., in prep). It is only applicable to proto-galactic halos prior to cosmic ionization due to its training dataset being the Rarepeak region in the \textit{Renaissance} Simulations that stops at $z = 15$.

The prime applications for such a model would be ones that do not include the effects of Pop III star formation and feedback, $\rm{H}_{2}$ chemistry, or do not resolve minihalos.  There are multiple reasons for not including these processes with a major reason being saving on computational expenses. Most heavy seeding models in simulations implant a BH once the halo crossing a mass threshold \citep{Schaye2015,Dave2019,Lovell2021,Vijayan2021}, which other models consider more stringent criteria such as low metallicity, temperature, and stellar component properties \citep[e.g.][]{Tremmel2017,Kaviraj2017}.  These models assume that it will have grown to the current $M_{\bullet}-\sigma$ relation by that point. Our results can only be applied to theoretical models that resolve the smallest atomic cooling halos, where we have restricted our analysis to halos with a mass within a factor of ($1 \pm 0.3$) of the atomic cooling limit (Equation \ref{eqn:acl}).

A probabilistic seeding model will provide a wider range of BH to stellar mass ratios. Because they are forming from proto-galactic gas in a low-mass starless halo, the initial star formation event may be affected heavily by early BH growth and feedback. Whether or not the presence of a central MBH affects the galactic properties enough to be observable remains to be seen.  Furthermore, while the initial $M_{\bullet}/M_{*}$ ratios may be overly massive, the evolution of the stellar component and BH will depend on the subsequent halo growth history and associated gas supply and ensuing star formation and feedback \citep{Sijacki2015,Scoggins2023,Dattathri2024,Guia2024,Pacucci2024,Shimizu2024,Sun2025}.

Lastly, it is not uncommon for these model galaxies to have an overly massive BH but they may only exist for a short period (tens of megayears) before the stellar component grows substantially \citep[e.g.][]{Scoggins2023}. JWST has detected these overly massive BH galaxies, and they may have been caught in this short phase of their evolution \citep{Bunker2023,Bogdan2024,Furtak2024,Juodzbalis2024,Maiolino2024a,Maiolino2024b,Marshall2024,Natarajan2024}. However, it is not irrefutable evidence for massive black hole formation, which would ideally catch the process in the act or shortly afterwards through unique observables associated with DCBH formation.

\subsection{Caveats}

Our work uses the \textit{Renaissance} Simulations for our statistical analysis of DCBH formation sites. Although it is a zoom-in simulation suite and has high resolution, its $29{,}000 \Msun$ dark matter resolution only allows it to resolve halos down to $3 \times 10^{6} \Msun$ with 100 particles. The authors have justified this in that they capture most Pop III star formation because LW radiation and streaming velocities can suppress Pop III star formation at the lowest minihalo mass scales ($10^{5}-10^{6} \Msun$). Nevertheless, the \textit{Renaissance} Simulations may be missing some Pop III star formation at these halo mass scales, whose feedback would be counterproductive to DCBH formation at the atomic cooling scale. This effect would induce an overestimate of DCBH formation sites in our model. On the other hand, we have restricted our formation sites to be nearly metal-free ($Z < 10^{-4} \Zsun$), which some recent works have shown that massive seeds may form in metal-enriched halos \citep{Regan2020b,Chon2021,Hirano2023}.  If this were the case, our results would underestimate the number of DCBH formation sites, most possibly making metallicity less of an important feature.

We have explored multiple methods to compare these two disparately sized populations and determine their feature importance. The feature rankings differ between the Mahalanobis and permutation methods and whether the most or least important feature was removed. While the individual feature rankings had more variations, their movements were mostly contained within the correlated groups. Although these variations may instill less confidence in our findings, the main outlier in the approaches is the Mahalanobis method, eliminating the least important features first. There is general agreement in the rankings for the individual and grouped features, both for the Mahalanobis (Section \ref{subsubsec:resmahal}) and Random Forest Classifier (Section \ref{subsubsec:resperm}) methods. They both recovered the halo mass and metallicity as being the most important, which were criteria for the DCBH candidates, and favored the central core quantities as the next most important quantities, which agrees with analytical expectations.

Lastly, the candidate DCBH halos are only candidates because the parsec-scale resolution of the \textit{Renaissance} Simulations is not sufficient to follow the central fragmentation. However, given that our ultra-high resolution re-simulations of three candidates resulted in very massive clumps and stars \citep{Wise2019,Regan2020c}, we are confident that a high fraction of our candidates will host massive black hole formation. We are currently following up on these 35 candidate halos in simulations similar to \citet{Wise2019} to determine their initial central object. In closing, our work quantifies the important halo features where massive black hole formation is most likely to occur. This information is useful for building subgrid models or identifying halos for targeted re-simulations to further investigate potential DCBH formation in more detail, searching for observable signatures that are unique to this process and the early universe.

\subsection{A Note on Effects of Selection Methods}
Our work makes certain assumptions on what classifies a halo as a candidate for direct collapse. Our selection follows the selection criteria of starless halos above the atomic cooling limit and is below a metallicity threshold of $10^{-4} \mathrm{Z}_\odot$. We explored the effects of our candidate selection methods on feature selection, specifically removing the metal-poor requirement as this feature comes out as important in all our selection methods.  Physically, this more relaxed selection criteria allows for halos that are externally enriched by nearby galaxies or enriched from Pop III star formation in their progenitors to be included.  Without any stars in these halos, prior work has shown that metal-poor ($Z \lesssim 10^{-3} \mathrm{Z}_\odot$) central gaseous clouds could still maintain a high enough infall rate without substantial fragmentation to support DCBH formation \citep{Chon2024}.  In systems with higher metallicites, fragmentation could lead to a dense stellar cluster that could proceed to form a massive black hole through stellar collisions \citep[e.g.][]{Portegies2004, Sakurai2017}.

Performing the feature selection methods with this new group of candidate halos, in which our sample grows to 78, we find that metallicity is still among the top features in all instances having a lowest ranked spot of four. Many of our top values remained consistently high such as halo mass, density, and radial mass flux. Additionally, bottom values like tidal field t$_1$ remained of little importance. However, in grouped rankings growth rates moved down in the rankings in favor of the halo exterior factors.

In performing recursive feature ranking most features stayed within a $\pm3$ range of its original ranking, however a few features notably changed substantially mostly within the lower to middle importance range. The same changes were not seen within the other ranking methods as such it likely has to do with the model. When the metallicity criteria is removed the Mahalanobis distance decreases from $\sim2.7$ to $\sim1.65$ a whole point decrease, showing the candidates are more statistically similar to the total population. Thus, including the metallicity criteria makes the probability of accurately training a model higher.

\section{Conclusion} \label{sec:con}
The aim of this paper is to investigate the properties of SMBH formation through direct collapse. Currently, three primary seeding mechanisms exist that develop into a BH in the early universe, light ($M_{\bullet} \lesssim 10^{2} \Msun$), intermediate ($M_{\bullet} \sim 10^{3} \Msun$), and heavy ($M_{\bullet} \sim 10^{4} \Msun$) seeds. The mechanism most likely to produce a SMBH in the early universe is the heavy or direct collapse mechanism which precedes a black hole through direct isothermal collapse. The majority of seeding models available use only mass and/or density to select DCBH host candidate halos. We explore the prominence of other halo features in determining candidacy. To do so we use the Rarepeak simulations from the \textit{Renaissance} suite to investigate potential features that indicate future DCBH formation. We examine a total of 35 candidates in this region and compare to $\sim\!4000$ non-candidates. Overall, we find that the trends support the importance of central halo properties over environmental properties.

\begin{enumerate}
    \itemsep0em 
    \item Our first main result demonstrates that DCBH host halos are significantly distant from non-candidates to be considered statistically separate populations, qualifying the creation of a subgrid model. This result follows from the high \rm{Z}-scores for multiple features and a Mahalanobis distance $>1$.
    \item Excepting properties used for candidacy selection, we identify the central density and radial mass influx (through both machine learning and statistical methods) as the features of most importance in determining capability of hosting a DCBH, with both features higher than average for candidates.
    \item Our results support the recent departure from the concept of a "Goldilocks zone" for galaxy-host halo distance as well as the minimal affect of Lyman-Werner flux on candidacy -- suggesting a different cooling suppressant.
    \item In halos with similar total mass, candidate halos contain a much denser core and fewer filaments stretching outside the central halo, making halo mass insufficient to determine candidacy without the inclusion of halo structure and history. This is a result of the non-candidate halos experiencing prior star formation leading to fragmentation, whereas candidates do not undergo star formation so fragmentation does not occur.
    \item Based on statistical testing and the case study, we can establish that the $\rm{H}_{2}$ fraction is negligible in both candidate and non-candidate halos with minimal to no recent star formation; thus, the starless condition on candidate halos already selects for low $\rm{H}_{2}$, removing the necessity of this feature.
    \item Based on their growth history, candidate halos tend to begin at a low mass but gain more over time than non-candidates, having increased in mass by approximately a factor of $12$ compared to the non-candidates doubling in mass over the final $130 \, \rm{Myr}$. This rapid growth may explain the suppression of $\rm{H}_{2}$ in these halos.
\end{enumerate}

In summary, seeding models should focus on halo density and core growth features to determine DCBH candidacy. We again make note that the \textit{Renaissance} simulations do not track black hole formation through collapse, instead we choose candidates based on atomic cooling halo requirements. In the future, we plan to run zoom-in halo simulations to track collapse and confirm if candidates will produce a DCBH(s).

As JWST continues to focus on the evolution of the early Universe it becomes more prevalent to provide a framework for halos capable of direct collapse to identify key observables of massive black hole formation. To fully actualize a framework for assessing the probability a halo will form a SMBH through direct collapse, we will extend this work by training a machine learning model to inform a DCBH formation subgrid model to be applicable in numerous cosmological simulation codes.


\section*{Acknowledgments}

We thank an anonymous referee whose insightful feedback strengthened the overall quality of the paper. This work was supported by NSF grant AST-2108020 and NASA grant 80NSSC21K1053.  EM acknowledges support from the President's Undergraduate Research Award from Georgia Tech.  The analysis for this work was performed on the Phoenix cluster within Georgia Tech's Partnership for an Advanced Computing Environment (PACE).  The \textit{Renaissance} Simulations were performed on Blue Waters operated by the National Center for Supercomputing Applications (NCSA) with PRAC allocation support by the NSF (awards ACI-0832662, ACI-1238993, and ACI-1514580).   This research is part of the Blue Waters sustained-petascale computing project, which is supported by the NSF (awards OCI-0725070, ACI-1238993) and the state of Illinois. Blue Waters is a joint effort of the University of Illinois at Urbana-Champaign and its NCSA.  The freely available astrophysical analysis code \texttt{yt} \citep{Turk2011} and plotting library matplotlib \citep{Hunter2007} were used to construct numerous plots within this paper. Computations described in this work were performed using the publicly-available \texttt{\textsc{Enzo}} code, which is the product of a collaborative effort of many independent scientists from numerous institutions around the world.

\bibliography{references}{}
\bibliographystyle{aasjournal}

\end{document}